\documentclass[aps,prb,twocolumn,groupaddress]{revtex4}
\usepackage{graphicx}
\usepackage{dcolumn}
\usepackage{bm}
\usepackage{color}
\begin{document}
%
%
\title{Impact of bicritical fluctuation on magnetocaloric phenomena in perovskite manganites}
%
\author{H. Sakai$^1$, Y. Taguchi$^1$, and Y. Tokura$^{1, 2, 3}$}
\affiliation{$^1$Cross-Correlated Materials Research Group (CMRG), ASI, RIKEN, Wako 351-0198, Japan\\
$^2$Multiferroics Project, ERATO, Japan Science and Technology Agency (JST), Tokyo 113-8656, Japan\\
$^3$Department of Applied Physics, University of Tokyo, Tokyo 113-8656, Japan}
%
\begin{abstract}
Variation of magnetocaloric (MC) effects has been systematically investigated for colossal magnetoresistive (CMR) manganites $R_{0.6}$Sr$_{0.4}$MnO$_{3}$ ($R$=La-Gd).
As the one-electron bandwidth is reduced, the temperature profile of MC effect, $i$.$e$. field-induced entropy change, exhibits a steeper drop below the ferromagnetic transition temperature due to its first-order nature promoted by a competing charge-orbital ordering instability.
For these small-bandwidth systems adjacent to the metal-insulator phase boundary, a rectangular-shaped profile for the entropy change emerges with an anomalously wide temperature range and a considerable magnitude.
Model calculations have indicated that the fluctuation enhanced in the phase-competing region has a strong impact on such MC features, which can be extensively controlled by the chemical composition.
\end{abstract}
%
\maketitle
%
A magnetocaloric (MC) effect refers to the isothermal entropy change, $\Delta S$, and the adiabatic temperature change, $\Delta T_{\rm ad}$, induced by applying (or removing) a magnetic field to the materials.
Since it constitutes the basis of the magnetic refrigeration with environmental friendliness and high efficiency, the quest for giant MC materials is currently of great interest.\cite{Gschneider2005review} 
It was recently found that systems undergoing a first-order magnetic transition show a large $\Delta S$ at the transition temperature, $T_{\rm C}$.
Alloyed metals with $T_{\rm C}$ near room temperature were hence the main subject of research, aiming at the application to refrigerators and air-conditioners.\cite{Pecharsky1997PRLa,Wada2001APLa,Fujieda2002APLa}
In addition, the low-temperature refrigeration such as hydrogen liquefaction now requires the wide-range refrigeration ability down to $\sim$20 K.
In such contexts, colossal magnetoresistive (CMR) manganites, where $T_{\rm C}$ of ferromagnetic metallic (FM) phase can be widely tuned by changing the one-electron bandwidth, can be promising MC materials.
Although various large-bandwidth manganites have been intensively investigated so far,\cite{Phan2007JMMMa} systematic studies focusing on the narrow-bandwidth systems with low $T_{\rm C}$ are quite rare.\cite{Sarkar2008APLa,Rebello2008APLa}
In the latter case, as a result of phase competition with a charge-orbital-ordered (CO/OO) insulator, sharp FM transitions are observed as the CMR phenomena,\cite{Tokura2006review} where substantial $\Delta S$ may be generated by a weak field.
%
\par
\begin{figure}
\includegraphics[width=7cm]{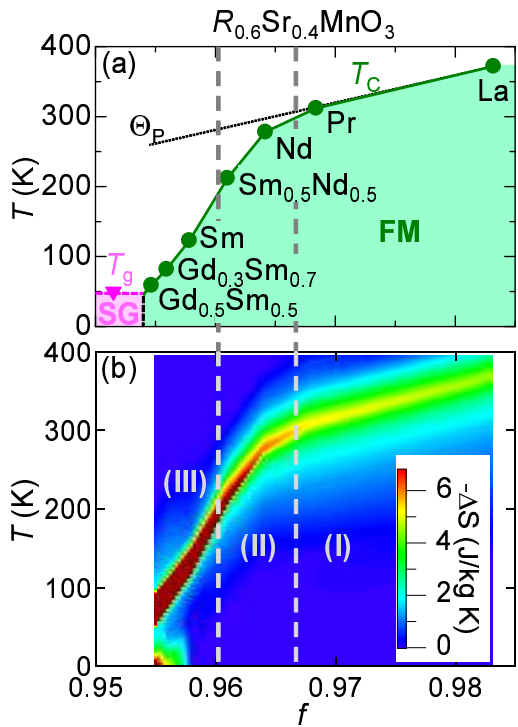}
\caption{\label{fig:contour}(Color online) (a) Phase diagram for $R_{0.6}$Sr$_{0.4}$MnO$_3$ ($R$=La-Gd) as a function of  a tolerance factor $f$, which is defined in the text. The ferromagnetic metal and spin-glass insulator are denoted by FM and SG, respectively. The ferromagnetic transition temperature $T_{\rm C}$, determined as the peak position of $-\Delta S$, is represented by closed circles, while the SG transition temperature $T_{\rm g}$ by closed triangles. (b) Contour plot of $-\Delta S$ for $H$=6 T on the $f$-$T$ plane.}
\end{figure}

%
The purpose of this Letter is to reveal the comprehensive feature and the underlying new physics for the MC effect of the CMR manganites.
As an ideal arena for this, we have adopted a series of single crystals of $R_{0.6}$Sr$_{0.4}$MnO$_{3}$ ($R$=La-Gd), which exhibit the wide-ranging $T_{\rm C}$ from 375 K to 59 K by decreasing the ionic radius of $R$ site, or the bandwidth [Fig. \ref{fig:contour}(a)].
The magnitude of the bandwidth is determined by a tolerance factor $f$, defined such that $f\!=\!(r_{\rm A}\!+\!r_{\rm O})/\sqrt{2}(r_{\rm Mn}\!+\!r_{\rm O})$, with $r_{\rm A}$, $r_{\rm Mn}$, and $r_{\rm O}$ being the (averaged) ionic radii of the perovskite A- and Mn-site cations and oxygen, respectively.\cite{Torrance1992PRBa,Imada1998RMP}
Even in the most reduced-bandwidth regime, no long-range CO/OO phase is realized due to the quenched-disorder effect arising from large mismatch in ionic size of $R$ and Sr, and instead a spin glass (SG) phase shows up with only short-range CO/OO correlation.\cite{Tomioka2003PRBa}
We have revealed a systematic change in MC properties throughout the phase diagram and, in particular, an anomalous MC effect characterized by a wide temperature range exhibiting substantial $\Delta S$.
Detailed comparison with model calculations indicates that the origin is ascribed to the gigantic phase fluctuation evolving near the phase boundary to the SG phase.
%
\par
%
Single crystals of $R_{0.6}$Sr$_{0.4}$MnO$_3$ ($R$=La-Gd) were grown by a floating zone method.\cite{Tomioka1997APLa, Sakai2008JPSJa}
Powder x-ray diffraction exhibited that the obtained crystals are of single phase.
The magnetization was measured with a superconducting quantum interference device (Quantum Design).
The specific heat measurement was performed by a standard heat relaxation method, using Physical Property Measurement System (Quantum Design).
%
\par
\begin{figure}
\includegraphics[width=8cm]{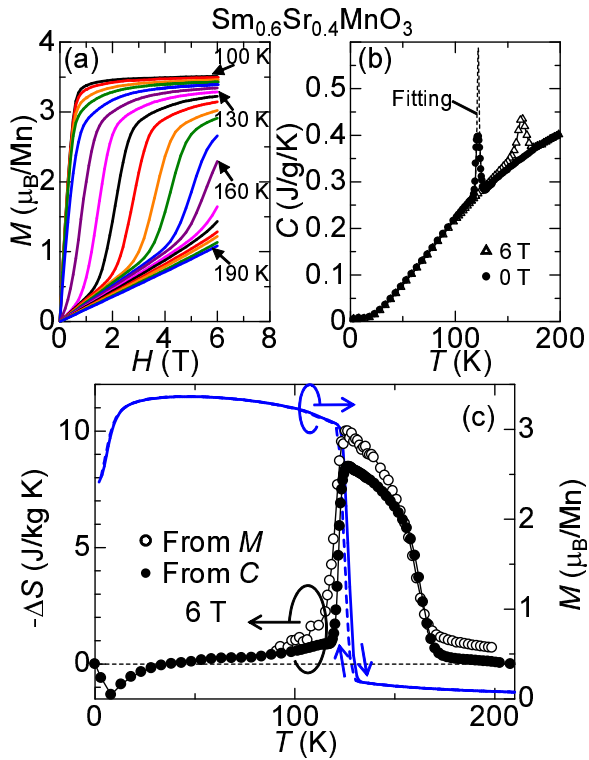}
\caption{\label{fig:SSMO}(Color online) Magnetocaloric properties for $R$=Sm. (a) Isothermal magnetization ($M$) versus magnetic field ($H$) up to around $T_{\rm C}$, taken with the temperature ($T$) interval of 2 K (not all shown here) in the $H$- and $T$-decreasing runs. (b) Temperature profile of specific heat ($C$) at 0 T and 6 T. The dashed line is the Lorentzian fitting to the latent-heat peak. (c) Temperature dependence of entropy change for $H$=6 T, calculated from both $M$ and $C$ data. The corresponding $M$ at 0.5 T is also shown.}
\end{figure}

%
We first focus on the $R$=Sm compound, which exhibits a first-order FM transition around 123 K [Fig. \ref{fig:SSMO}(c)], as a typical example.
We can indirectly obtain $\Delta S$ by utilizing the Maxwell relation: $\Delta S(H, T)\!=\!\int^{H}_{0}(\partial M/\partial T)_{H}\,{\rm d}H$, where $H$, $T$, and $M$ are applied field, temperature, and magnetization, respectively.
Using all the discrete $M$-$H$ data shown in Fig. \ref{fig:SSMO}(a), we numerically calculated the $\Delta S$ values [open circles in Fig. \ref{fig:SSMO}(c)].
As an alternative, measurements of the specific heat $C(H, T)$ at $H$=0 T and 6 T [Fig. \ref{fig:SSMO}(b)] can also figure out $\Delta S$ based on the thermodynamic law: $\Delta S(H, T)\!=\!\int^{T}_{0}[C(H, T)-C(0, T)]/T\,{\rm d}T$.
Because of the difficulty in measuring the sharp peak stemming from the latent-heat contribution at 0 T by the relaxation method, we interpolated the data by a Lorentzian function following the literature,\cite{Gordon2001PRBa} when performing the numerical $T$ integration.
Thus calculated $\Delta S$ values [closed circles in Fig. \ref{fig:SSMO}(c)] are in good agreement with those from the $M$ data.
This consistency between the magnetic and calorimetric measurements ensures the validity of relying on the Maxwell relation in the present systems, despite their first-order nature of the transition.\cite{Liu2007APLa}
In the following, we have estimated $\Delta S$ from the $M$ data.
%
\par
%
\begin{figure}
\includegraphics[width=7cm]{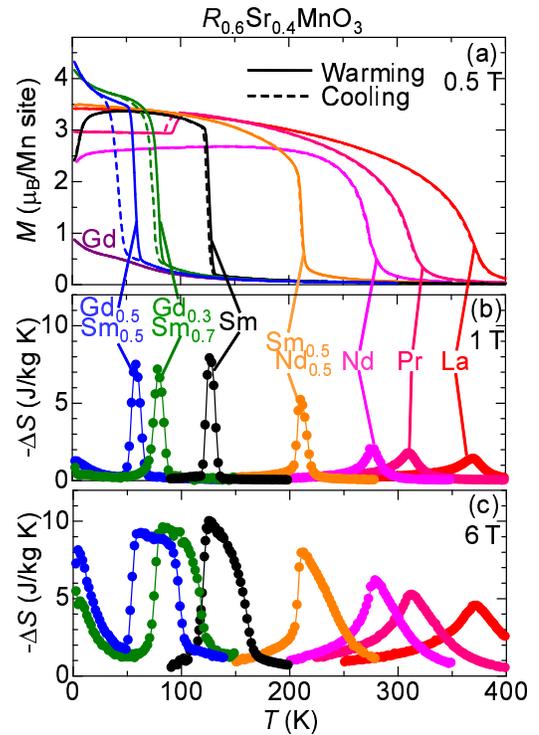}
\caption{\label{fig:RSMO}(Color online) Temperature profiles of $M$ at 0.5 T (a) and $\Delta S$ for $H$=1 T (b) and 6 T (c) for the single crystals of $R_{0.6}$Sr$_{0.4}$MnO$_{3}$ ($R$=La-Gd). The solid and dashed lines in (a) correspond to warming and cooling runs, respectively. The $\Delta S$ data was obtained in the $H$ and $T$-decreasing runs.}
\end{figure}

%
We show the temperature dependence of $M$ for $R_{0.6}$Sr$_{0.4}$MnO$_{3}$ ($R$=La-Gd) at 0.5 T in Fig. \ref{fig:RSMO}(a).
For the $R$=La crystal with the largest bandwidth, a simple second-order FM transition was discerned at $T_{\rm C}$ as high as 375 K.
With decreasing the bandwidth, $T_{\rm C}$ systematically decreases and the FM transition becomes steeper.
Judging from the temperature hysteresis at the transition, the critical end point of the first-order transition seems to locate in the vicinity of $R$=Sm$_{0.5}$Nd$_{0.5}$.\cite{Demko2008PRLa}
In the systems with smaller $f$, such as $R$=Gd$_{1-x}$Sm$_{x}$ ($0.5\!\le\! x \!\le\!1$), the apparent first-order FM transition was observed, except for the $R$=Gd crystal (SG insulator).
%
%
\par
%
The resultant curves of $-\Delta S$ are shown in Figs. \ref{fig:RSMO}(b) and (c) for $H$=1 T and 6 T, respectively.
The temperature range of the significant MC effect can be widely controlled from 50 K to 400 K by changing the $R$ species.
In the large-bandwidth systems ($R$=La and Pr), $-\Delta S$ exhibits a broad peak at $T_{\rm C}$.
Both its height and width steadily increase with increasing the magnitude of $H$, characteristic of the second-order transition.
When the bandwidth of the system is reduced, the shape of $-\Delta S$ becomes sharper with a sudden drop below $T_{\rm C}$.
Accordingly, the peak value becomes larger and reaches $\sim$10 J/kg K ($\sim$18\% of the total magnetic entropy) for $H$=6 T for $R$=Sm, although it saturates for $R$=Gd$_{1-x}$Sm$_{x}$ ($0.5\!\le\! x \!\le\!1$).
There, the height of the $-\Delta S$ peak is not sensitive to $H$ above 1 T, whereas its width increases nearly linearly with $H$, as is typical of the first-order transition.
Noteworthy is that their peak shape changes to the rectangular one with an anomalously wide plateau temperature region (e.g., $\sim$50 K for $R$=Gd$_{0.5}$Sm$_{0.5}$ for $H$=6 T), as compared with the case of rare-earth alloys showing the first-order transition at similar $T_{\rm C}$.\cite{Wada1999Cryo}
A gradual but large increase in $-\Delta S$ below 40 K is probably due to the contribution from the Gd spins.
%
\par
%
To provide a perspective of the MC properties for $R_{0.6}$Sr$_{0.4}$MnO$_3$ ($R$=La-Gd), we display in Fig. \ref{fig:contour}(b) a contour plot of $-\Delta S$ for $H$=6 T on the entire $f$-$T$ plane.
Contour values are determined by interpolation based on the data shown in Fig. \ref{fig:RSMO}(c).
Depending on the behavior of $\Delta S$, we roughly classify their MC effect into three regions in the $f$-$T$ plane.
In the region (I) corresponding to $R$=La-Pr with the second-order FM transition, $-\Delta S$ exhibits a broad peak around $T_{\rm C}$, forming a yellow-colored (online) hump.
In the intermediate-bandwidth region (II), such as $R$=Nd-Sm$_{0.5}$Nd$_{0.5}$, $T_{\rm C}$ begins to rapidly decrease with decreasing the bandwidth due to the enhanced antiferromagnetic CO/OO correlation.\cite{Saitoh1999PRBa}
Furthermore, the FM transition approaches the first-order one accompanied by a sharp drop in $-\Delta S$ below $T_{\rm C}$, which results in the evolution of the red-colored (online) sharp ridge.
In the region (III) corresponding to $R$=Gd$_{1-x}$Sm$_{x}$ ($0.5\!\le\!x\!\le\!1$) with the narrowest-bandwidth, $T_{\rm C}$ is markedly suppressed as the SG phase is approached.
These crystals undergo a first-order transition even at high fields, leading to a steep rise in $-\Delta S$ above $T_{\rm C}$.
The resultant rectangular-shaped $-\Delta S$ is highlighted as the red-colored (online) plateau structure below $\sim$150 K.
In each case, we hereinafter make a detailed comparison between the model calculation and the observed data.
%
%
\par
\begin{figure}
\includegraphics[width=6cm]{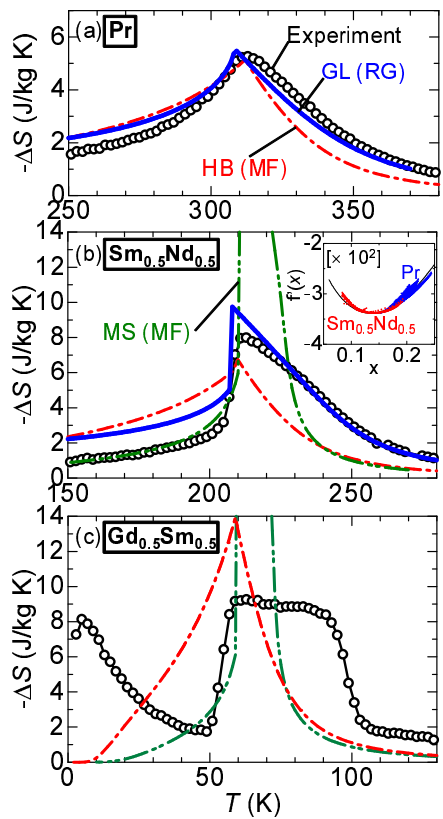}
\caption{\label{fig:calculation}(Color online) Temperature profiles of $\Delta S$ around $T_{\rm C}$ for $H$=6 T for $R$=Pr (a), Sm$_{0.5}$Nd$_{0.5}$ (b), and Gd$_{0.5}$Sm$_{0.5}$ (c). Results of the mean-field (MF) calculation using the Heisenberg (HB) model and that with magnetostriction effects (MS) are shown by dashed-dotted and dashed-two-dotted lines, respectively, while those calculated using the Ginzburg-Landau (GL) theory based on the renormalization group (RG) analysis shown by solid lines. Experimental data correspond to open circles. Inset to (b) exhibits the scaling plot for $R$=Pr and Sm$_{0.5}$Nd$_{0.5}$ with the curve of $f'(x)\!=\!\frac{x}{4}(\ln x\!+\!1)$.}
\end{figure}

%
In the region (I), the mean-field calculation of $\Delta S$ using the Heisenberg model of spin $s$=1.8 with the equivalent $T_{\rm C}$ can almost reproduce the experimental values, as exemplified for $R$=Pr in Fig. \ref{fig:calculation}(a).
Because of the large spin value and the weak CO/OO instability, the mean-field approximation works well in this regime.
Nevertheless, a small deviation is discerned particularly above $T_{\rm C}$, which can be improved by the theory taking account of the bicritical fluctuation ($vide$ $infra$).
%
\par
%
As shown for $R$=Sm$_{0.5}$Nd$_{0.5}$ in Fig. \ref{fig:calculation}(b), the Heisenberg model can no longer reproduce the sharp drop in $-\Delta S$ below $T_{\rm C}$ in the region (II).
To compare this behavior with the first-order transition model, we took account of the magnetostriction effect in the mean-field Heisenberg Hamiltonian:\cite{Bean1962PRa, Ranke2004JMMMa, Zou2008EPLa} $\mathcal{H}_{\rm MF}\!=\![-2zJ(\varepsilon)\langle s\rangle\!-\!g\mu_{\rm B}H]\sum_{i}s_{i}\!+\!\sum_{\langle i,j\rangle}[2J(\varepsilon)\langle s\rangle^{2}\!+\!(\kappa/2)(1\!-\!\varepsilon)^{2}]$, where $\varepsilon$ is the amount of the dimensionless lattice strain, $z$ the number of the nearest neighbors, $\langle s\rangle$ the thermodynamic average of the spin, $s_{i}$ the spin operator at the $i$ site, $g$ the Land\'{e} factor, $\mu_{\rm B}$ the Bohr magneton, and $\kappa$ the elastic constant.
The magnetostrictive exchange energy is given by $J(\varepsilon)\!=\!J_{0}\!-\!\lambda\varepsilon$, where $J_{0}$ is determined as $2zJ_{0}\!=\!3k_{\rm B}\Theta_{\rm P}/s(s+1)$ ($k_{\rm B}$ is the Boltzmann constant) and $\lambda$ is a positive coefficient.
$\Theta_{\rm P}$ corresponds to the FM transition temperature, provided that the one-electron bandwidth is simply reduced without the CO/OO instability [see Fig. \ref{fig:contour}(a)].
The elastic energy term $\frac{\kappa}{2}(1-\varepsilon)^{2}$ effectively includes the sum of the energy loss due to the lattice strain and the energy gain due to the Jahn-Teller effect, arising from the CO/OO correlation.
Because $\varepsilon\!=\!1$ (for $H$=0 T) above $T_{C}$ so as to minimize this term, the reduced ferromagnetic exchange coupling $J_{0}\!-\!\lambda\varepsilon$ results in $T_{\rm C}$ lower than $\Theta_{\rm P}$, which mimics the short-range CO/OO correlation evolving above $T_{\rm C}$.
After calculating $\langle s\rangle$ and $\varepsilon$ by minimizing the free energy, we determined $\lambda$ and $\kappa$ so that $T_{\rm C}$ and the temperature profile of $\Delta S$ are most consistent with the experimental results.
The result fitted for $R$=Sm$_{0.5}$Nd$_{0.5}$ is shown by a dashed-two-dotted line in Fig. \ref{fig:calculation}(b).
Whereas it reproduces the $-\Delta S$ drop below $T_{\rm C}$, its peak height is $\sim$2 times as high as the experimental value.
Furthermore, the predicted behavior above $T_{\rm C}$ differs significantly from the experimental results.
%
\par
%
To correctly take into consideration the effect of fluctuations enhanced in the phase-competing region (II) beyond the mean-field scheme, we now deal with the phenomenological Ginzburg-Landau model based on the renormalization group analysis.\cite{Murakami2003PRLa}
This method not only successfully explains the first-order FM transition around the bicritical point for CMR manganites, but also derives the universal scaling relation between $H/M$ and $M^{2}$ (Arrot plot), expressed as $H/M\!=\!r_{M}\!+\!2g_{\rho M}N_{\rho}f'(r_{\rho}+g_{\rho M}M^{2})$, where $r_{M}$, $r_{\rho}$, $N_{\rho}$, and $g_{\rho M}$ are scaling parameters, $\rho$ the order parameter of charge ordering, and $f'(x)\!=\!\frac{x}{4}(\ln x\!+\!1)$.
Following the literature,\cite{Murakami2003PRLa} we performed the scaling plot of the $M$-$H$ data for $R$=Pr--Sm$_{0.5}$Nd$_{0.5}$ by regarding $f$ as a parameter controlling the bandwidth.
Note here that $r_{M}$ and $r_{\rho}$ are expanded with respect to $T$ and $f$ near the bicritical point: $r_{i}\!=\!c_{iT}\Delta T\!+\!c_{if}\Delta f$ ($i\!=\!M$, $\rho$), where $c_{iT}$ and $c_{if}$ are constants independent of $R$.
The results of the scaling plot for $R$=Pr--Sm$_{0.5}$Nd$_{0.5}$ are shown in the inset to Fig. \ref{fig:calculation}(b).
All the $M$-$H$ data fall into one curve with good accuracy, reflecting that the bicritical fluctuation is enhanced in this regime.
Using the obtained parameters, we calculated $\Delta S$ for $R$=Sm$_{0.5}$Nd$_{0.5}$ [Fig. \ref{fig:calculation}(b)].
The result well reproduces the broad tail above $T_{\rm C}$ as well as the steep drop at $T_{\rm C}$, which the mean-field calculation fails.
The enhanced fluctuation makes the system sensitive to a magnetic field, which may effectively broaden the peak width of $-\Delta S$.
Also for $R$=Pr, the renormalization group analysis works better than the mean-field one [Fig. \ref{fig:calculation}(a)], indicating some influence of the fluctuation even in the region (I).
%
%
\par
%
The plateau structure of $-\Delta S$ observed in the region (III) cannot be reproduced by the mean-field calculation in either the Heisenberg model nor that with magnetostriction, as shown for $R$=Gd$_{0.5}$Sm$_{0.5}$ in Fig. \ref{fig:calculation}(c).
(We are not concerned here with the broad peak at the lowest temperature, stemming from the Gd spins.)
Although the latter model can reproduce a rectangular shape, its width is less than $1/3$ of that observed and its height is about 3 times.
This anomalously wide plateau might be also caused by the fluctuation further developed in the low-$T_{\rm C}$ systems.
As described in the literature,\cite{Murakami2003PRLa} however, the scaling relation cannot be applied to the strong-fluctuation regime accompanied by the distinct first-order transition like the region (III).
In this regime, quenched disorder arising from the A-site solid solution completely suppresses the long-range CO/OO phase.
It leads to the short-range CO/OO state evolving above $T_{\rm C}$,\cite{Tomioka2003PRBa} which tends to promote the first-order nature of the FM transition.
To correctly deal with their MC phenomena, we hence need a more elaborate model which takes account of the strong quenched disorder in addition to the phase competition.
%
\par
%
From a practical viewpoint, this series of manganites may be suitable as a magnetic refrigerant especially below $\sim$200 K.
There, they indicate the good performance: a refrigeration range of as wide as $\sim$50 K with substantial heat extraction ($-\Delta S\!\sim\!10$ J/kg K) for $H$=6 T, achieving the cooling capacity\cite{Gschneider2005review,Phan2007JMMMa} of $\sim$500 J/kg comparable with that of the giant MC materials.
Furthermore, the plateau-like shape of the $-\Delta S$ peak is suitable for the Ericsson cycle, where the constant $\Delta S$ as a function of temperature is required within the cooling region.\cite{Hashimoto1987JAPa, Korte1998PRBa}
%
\par
%
In conclusion, we have investigated the overall feature of the MC effects for a series of CMR manganites $R_{0.6}$Sr$_{0.4}$MnO$_{3}$ ($R$=La-Gd) with the controlled bandwidth.
With decreasing the bandwidth (or $T_{\rm C}$), the $-\Delta S$ peak becomes larger, changing to a rectangular shape, due to the first-order nature of the transition induced by the competition with the CO/OO instability.
By performing various model calculations, we have attributed the observed unusually wide temperature range of their MC effect to the gigantic fluctuation in the phase-competing regime.
Thus, the electron correlation can strongly affect (or sometimes improve) the MC features, giving a new clue to exploring giant MC materials.
%
\par
%
We thank Y. Onose and H. Katsura for fruitful discussions.
This study was partly supported by Grant-in-Aid from Special Postdoctoral Researchers Program and Incentive Research Grant in RIKEN.
%

%

\end{document}